\documentclass[12pt]{article}
\usepackage{amsfonts}
\usepackage[english]{babel}
\usepackage[T1]{fontenc}
\usepackage{amsthm,amsmath,amsfonts,amssymb,epsfig,latexsym}
\usepackage{bm}% bold math
\usepackage{helvet}         % selects Helvetica as sans-serif font
\usepackage{courier}        % selects Courier as typewriter font
\usepackage{multicol}        % used for the two-column index
\makeindex             % used for the subject index
\newcommand{\eg}{\emph{e.g. }}

\newcommand{\RR}{\mathbb{R}}
\newcommand{\NN}{\mathbb{N}}

\newcommand{\ff}{\hat f}

\newcommand{\fg}{\hat g}
\newcommand{\fu}{\hat u}

\newcommand{\R}{\mathbb R}
\def\be#1\ee{\begin{equation}#1\end{equation}}
\newcommand{\fer}[1]{(\ref{#1})}

\newcommand{\bq}{\begin{equation}}
\newcommand{\eq}{\end{equation}}

\def\bqa{\begin{eqnarray}}
\def\eqa{\end{eqnarray}}

\def\e{\epsilon}

\def\t{\tau}

\newcommand{\bd}{\begin{displaymath}}
\newcommand{\ed}{\end{displaymath}}
\newcommand{\ba}{\begin{eqnarray}}
\newcommand{\ea}{\end{eqnarray}}

\def\ff{\widehat f}

\def\N{\mathbb{N}}
\def\R {{I \!\! R}}

\begin{document}
%\begin{frontmatter}

\title{Kinetic and mean field description  of Gibrat's law}

\author{G. Toscani\thanks{Department of Mathematics, University of Pavia, via Ferrata 1, 27100 Pavia, Italy.
\texttt{giuseppe.toscani@unipv.it}} }
\date{}

\maketitle
\noindent
{\bf Abstract:} \small{We introduce and analyze a linear kinetic model that describes the evolution of the probability density of the number of firms in a society, in which the microscopic rate of change obeys to the so-called law of proportional effect proposed by Gibrat \cite{Gib1,Gib2}. Despite its apparent simplicity, the possible mean field limits of the kinetic model are varied. In some cases, the asymptotic limit can be described by a first-order partial differential equation. In other cases, the mean field equation is a linear diffusion with a non constant diffusion coefficient that  models also the geometric Brownian motion \cite{Oks} and can be studied analytically. In this case, it is shown that the large-time behavior of the solution is represented, for a large class of initial data, by a lognormal distribution with constant mean value and variance increasing exponentially in time at a precise rate. The relationship between the kinetic and the diffusion models allow to introduce an easy-to-implement expression for computing the Fourier transform of the lognormal distribution.}
\vskip 5mm

{\bf Keywords}
{Kinetic models; Gibrat's law; linear diffusion equations; large-time behavior;  lognormal distribution; Wild sums.}

%\end{frontmatter}

\section{Introduction}
The  agent-based models constitute a broad class of models which have
been recently introduced to describe various socio-economic phenomena of western societies \cite{NPT, PT13, Voi}. The mathematical modelling showed a great expansion especially in the past fifteen years  \cite{Ch02, ChaCha00, ChChMa04,
CYC, ChChSt05, CoPaTo05, DY00}. This relatively new research field
borrows several methods and tools from classical statistical
physics, where the macroscopic emergent behavior arises from relatively
simple rules as a consequence of microscopic interactions among a huge
number of agents \cite{NPT, PT13}. 

Kinetic models are often the building block. These models can be derived by resorting to well-known tools of
classical kinetic theory of gases \cite{CoPaTo05,DMT, DMT2, MaTo07,
Sl04}, where Boltzmann-like equation for Maxwell-type molecules play the relevant rule \cite{Cer, PT13}.  

Among the various interactions models that can be studied by this powerful methodology, one of the simplest ones is certainly the Gibrat's law for firm growth \cite{Gib1,Gib2}. Gibrat formulated the law of proportionate effect for growth rate to justify  the observed distributed distribution of firms. The law of proportionate effect states that the expected increment of a firm's size in a fixed period of time is proportional to the size of the firm at the beginning of the period.
Denoting by $x(\t)$ the size of a firm at a time $\t\ge0$, the postulate is expressed as
 \be\label{Gib}
 x(\t +1) = x(\t) + \eta(\t) x(\t),
 \ee
where $\eta(\t)$ is a random number independent of $x(\t)$, and $\eta(\t)$ is independent of $\eta(\t+k)$ for any natural number  $k$, and there are no interactions between firms. 

After a sufficiently long sequence of increments, since Gibrat's law implies that
 \[
 x(n) = x(0)(1+\eta(1))(1+\eta(2))\cdots (1+\eta(n)),
 \]
$\log x(n)$ follows a random walk. Therefore, the growth rate predicted by Gibrat's law is lognormally distributed with mean and variance linked to the mean and variance of $\eta(\cdot)$. The validity of Gibrat's law has been investigated by many authors, often with a critical viewpoint \cite{Gal, Mitz}.

Despite its simplicity, or maybe in reason of this, continuous kinetic models based on Gibrat's law seem to have not yet been derived in a rigorous way. Of course, the  rate of change expressed by \fer{Gib} appears as part of the microscopic binary interaction between agents in kinetic models for wealth distribution, like in the model proposed by the author with Cordier and Pareschi  \cite{CoPaTo05}, where the term $\eta(\t)x(\t)$ plays the role of the  risk in an economic trade in which $x(\t)$ denotes the wealth of the trader at time $\t$. Also, a law similar to \fer{Gib} appears in the pure gambling model studied in \cite{BaTo} to investigate the possibility to generate Pareto tails by conservative-in-the-mean interactions. 

In  this paper we aim to study both kinetic and mean field models generated by interactions of type \fer{Gib}. Depending on the properties of the random variable $\eta$, various limiting behaviors appear, that, while maintaining the main properties (conservation of the mean number of firms, growth of higher moments, etc.) exhibit completely different asymptotic behaviors. Among others,  we will show that Gibrat's law can be described in terms of the mean field equation 
 \be\label{gi-m}
 \frac{\partial u}{\partial t} = \frac\sigma 2\, \frac{\partial^2}{\partial x^2}(x^2 u),
 \ee
where $\sigma >0$ is a fixed constant. Equation \fer{gi-m} contains in fact  the main effects of Gibrat's law \fer{Gib} when the random variable $\eta$ produces small symmetric effects. The linear diffusion equation \fer{gi-m} allows to describe the evolution in time of the density $u=u(x,t)$ of the size $x \ge 0$ of firms, given their distribution $u_0(x)$ at time $t =0$, as well as its asymptotic behavior.  This equation is well-known to people working in probability theory and finance, since it describes a particular version of the geometric Brownian motion \cite{Oks}. It is noticeable that the solution of equation \fer{gi-m}  can be described analytically. Various phenomena are indeed described by  related equations. One of these phenomena has been recently investigated   by Iagar and S\'anchez \cite{IS1} in connection with the study of the heat equation in a nonhomogeneous medium with critical density.  There, at difference with the standard studies, they solve the  equation in the whole space. 

The interest in the rigorous relationship between the kinetic and diffusive models of Gibrat's law are also connected to the possibility to approximate the solution to the latter in terms of the solution to the former, which admits a very simple expression in terms of a Wild sum \cite{PT13, Wild} easy to obtain recursively. In particular this suggests a new way to look for numerical approximations to the Fourier transform of the lognormal distribution \cite{ALR}.

\section{The model}\label{model}

Let us consider a system composed of a huge number of agents  which are identified in terms of a certain characteristic, which can be modified by some universal interaction rule. If this characteristic is measured by a nonnegative number $x$, the aim of a kinetic model is to provide a continuous description for the evolution in time, denoted by $\t$,  of the density function $f(x,\t)$ of the $x$-variable consequent to interactions \cite{NPT, PT13}. 

Let us assume that the population of agents coincide with the list of firms. Then the precise  meaning of the density $f$ is the following. Given the list of firms to study, and a domain $D \subseteq  \R_+$, the integral
 \[
 \int_D f(x,\t) \, dx
 \]  
represents the percentage of firms with size included in $D$ at time $\t \ge 0$. A natural assumption is to normalize to one the density function, that is
\[
 \int_{\R_+} f(x,\t) \, dx =1
 \]  
for any time $\t\ge0$. According to  Gibrat's postulate \fer{Gib}, we will assume that the microscopic variation of the firm size is due to interactions with the external background, and it is proportional to the size itself. Consequently, given a firm of size $x$, its post-interaction size is given by
 \be\label{col}
 x^* = x + \eta x,
 \ee
where the random quantity  $\eta x$ represents the change in size  of the firm, proportional to the pre-interaction size $x$,  generated by the  presence of the  background. We will assume that the random variable $\eta$, takes values in a bounded set limited below by $-1$,  and it is of zero mean. The lower bound on $\eta$ will guarantee  that the post-interaction size $x^*$ will remain nonnegative. 

The study of the
time-evolution of the distribution of the size density produced by 
interactions  of type \fer{col}  can be obtained by
resorting to kinetic collision-like models \cite{Cer, PT13}, where the variation of the size density $f(x, \t)$  obeys to a
Boltzmann-like equation. This equation is usually written
in weak form. It corresponds to say that the solution $f(x,\t)$
satisfies, for all smooth functions $\varphi(x)$ (the observable quantities)
 \begin{equation}
  \label{kin-w}
 \frac{d}{d\t}\int_{\R_+}\varphi(x)\,f(x,\t)\,dx  = \lambda
  \Big \langle \int_{\R_+} \bigl( \varphi(x^*)-\varphi(x) \bigr) f(x,\t)
\,dx \Big \rangle.
 \end{equation}
As usual, $\langle \cdot \rangle$ represents mathematical expectation. Here expectation takes into account the presence of the random parameter $\eta$ in \fer{col}. The positive constant $\lambda$ measures the interaction frequency.

Clearly, the right-hand side of equation \fer{kin-w} represents a balance between the amount of firms that change their size from $x$ to $x^*$ (loss term with negative sign) and the amount of firms that move to the actual size $x$ from any other size $x^*$ (gain term with positive sign).

The choice $\varphi(x) = \exp\{-i\xi x\}$ shows in particular that the Fourier transform $\ff(\xi,\t)$ of the density, defined by
 \be\label{ft}
  \ff(\xi,\t) = \int_{\R_+} f(x,\t) e^{-i\xi x} \, dx
 \ee
satisfies the equation
 \be\label{kin-f}
 \frac{\partial}{\partial t} \ff(\xi,\t) = \lambda\left( \langle \ff((1+\eta)\xi, \t)\rangle -\ff(\xi,\t) \right).
 \ee
Equations \fer{kin-w} and \fer{kin-f}  are completed by assigning initial conditions $f(x, \t= 0) = f_0(x)$ (respectively $\ff(\xi, \t=0) = \ff_0(x)$). It is normally assumed that $f_0(x)$ is a probability density, $x \in \R_+$, and that the density possesses a certain number of bounded moments. In general, the physically relevant moment which is always assumed bounded is the average size (the mean value of the initial value).   

The main common properties of the kinetic equations \fer{kin-w} and \fer{kin-f} are easily derived by resorting to the form which is more adapted to the purpose. 

Existence and uniqueness of the solution to equation \fer{kin-w} can be obtained in a rather standard way, and for a large class of initial value densities, by resorting to classical methods of kinetic theory \cite{MaTo07}, which hold true also for bilinear kinetic equations. We refer to \cite{PT13} for a detailed description of these methods. In this paper, we will mainly concerned with asymptotic limit equations generated by \fer{kin-w}. For this reason, we need only to recover the main macroscopic features of the kinetic model. To this aim, let us first reckon the law of evolution of moments. By fixing $\varphi(x) = x^n$, $n \in \N_+$ we obtain from \fer{kin-w}
 \be\label{mome}
 \frac{d}{dt}\, m_n(\t) = \frac{d}{d\t}\int_{\R_+}x^n\,f(x,\t)\,dx  = \lambda
  \Big \langle (1+\eta)^n -1\Big \rangle m_n(\t).
 \ee
Since the random variable $\eta$ satisfies  $\langle \eta \rangle = 0$, the quantity $\lambda_n = \lambda\langle (1+\eta)^n -1 \rangle = 0$ when $n = 0, 1$. This shows conservation of mass, and, provided the first moment of the initial density is bounded, conservation of the average size. Then, any other moment which is initially bounded, increases exponentially at a rate $\lambda_n$. Since $\lambda_n$ depends on the values of the moments of the random variable $\eta$, the behavior of the solution is heavily dependent on $\eta$. 

Further insides on the evolution of the solution density $f(x, \t)$ require or a numerical approximation, or a simplification which can result from suitable limiting procedures, which are required to maintain the main macroscopic properties of the model. 
In what follows, we will examine various limit problems linked to equation \fer{kin-w}, which are generated by special choices of the random variable $\eta$.

\section{First-order continuous models}\label{first}

In the rest of the paper, without loss of generality, we will fix $\lambda =1$. It is clear that, by scaling time, we can always reduce the problem to this situation. 

Given a positive value $\epsilon\ll 1$, let us consider interactions of type \fer{col} produced through a random variable $\eta_\epsilon$ which can assume only the  values $\epsilon$ with probability $1-\epsilon$, and $\epsilon-1$ with probability $\epsilon$.  This choice corresponds to the situation in which there is a very high probability that the size of a firm could increase of a small amount, and a very small probability that the size of the firm would collapse.  The random variable $\eta_\e$ satisfies  $\langle \eta_\epsilon\rangle = 0$, and in addition  $\langle \eta_\epsilon^2 \rangle = \epsilon(1-\epsilon)$. Let $f_\epsilon(x,\t)$ denote the solution to equation \fer{kin-w} corresponding to the choice $\eta =\eta_\epsilon$. Then equation \fer{mome} for $n=2$ gives
 \[
 \frac{d}{d\t}\int_{\R_+}x^2\,f_\epsilon(x,\t)\,dx  = \epsilon(1-\epsilon) \int_{\R_+} x^2 f_\epsilon(x,\t)\, dx.
 \]
As $\epsilon \to 0$, the second moment of the solution tends to remain constant, thus loosing its typical property to increase with time. The property can be restored by scaling. Let us  choose $t = \epsilon \t $, and $f_\epsilon (x,\t) = g_\epsilon (x, t)$. Then, substituting in \fer{mome}, it is immediate to verify that  the second moment of $g_\epsilon$ satisfies the equation
  \[
 \frac{d}{dt}\int_{\R_+}x^2\,g_\epsilon(x,t)\,dx  = (1-\epsilon) \int_{\R_+} x^2 g_\epsilon(x,t)\, dx,
 \]
which ensures the standard exponential  growth independently of the value of the small parameter $\epsilon$. By means of this transformation, we can consequently investigate situations in which the interactions produce a very small variation of the firms size, simply by waiting enough time to maintain a strictly positive growth rate of the second moment.

Taking into account the simple expression of $\eta_\epsilon$, in Fourier transform equation \fer{kin-f} for $\fg_\epsilon(\xi, t)$ takes the form
\be\label{kin-g}
 \frac{\partial}{\partial t} \fg_\e(\xi,t) = 
\frac 1\epsilon \left( \fg_\e((1+\epsilon)\xi, t)(1-\epsilon) + \fg_\e(\epsilon \xi, t) \epsilon -\fg_\e(\xi,t) \right).
 \ee
Hence, letting $\e \to 0$, for any given time $t \ge 0$ the solution $\fg_\e(\xi, t)$ of equation \fer{kin-g} converges to $\fg(\xi,t)$, solution of the equation
 \be\label{k-g}
 \frac{\partial \fg(\xi,t)}{\partial t}  = 1 - \fg(\xi, t) + \xi \, \frac{\partial \fg(\xi,t)}{\partial \xi}  .
 \ee
The limit procedure can be made rigorous by resorting to Fourier-based metrics \cite{PT13}. We will present a proof for second-order models we will consider in Section \ref{diffu}. 
Equation \fer{k-g} contains most of the information relative to the choice of the random variable $\eta_\e$. In particular, we can easily extract from \fer{k-g} the growth of moments. 
 By taking the derivative with respect to $\xi$ in \fer{k-g} we obtain
 \[
 \frac{\partial \fg'(\xi,t)}{\partial t}  =  \xi\, \fg''(\xi, t). 
 \]
Hence, it is a simple exercise to obtain recursively that the subsequent derivatives with respect to $\xi$  of $\fg(\xi,t)$, say $\fg^{(n)}(\xi,t)$, $n > 1$,  satisfy the equations
 \[
 \frac{\partial \fg^{(n)}(\xi,t)}{\partial t}  = (n-1)\fg^{(n)}(\xi, t) + \xi\, \fg^{(n+1)}(\xi, t), 
 \]
 which imply, for $n \ge 1$
 \[
  \frac{\partial \fg^{(n)}(0,t)}{\partial t}  = (n-1) \,\fg^{(n)}(0, t),
 \]
or, what is the same
 \be\label{mn}
\frac{d}{dt}\int_{\R_+}x^n\,g(x,t)\,dx  = (n-1) \int_{\R_+}x^n\,g(x,t)\,dx.
 \ee
Consequently, equation \fer{k-g} preserves mass and average size, while the moments of order $n$ increase exponentially at rate $n-1$. 

Equation \fer{k-g} is explicitly solvable. Along characteristics,  one shows that \fer{k-g} is equivalent to 
 \be\label{cha}
 \frac d{dt} \fg(\xi e^{-t},t) =  1- \fg(\xi e^{-t},t),
 \ee
which can be integrated by separation of variables. Hence, if $\fg_0(\xi)$ denotes the Fourier transform of the initial density, the explicit solution of equation \fer{k-g} reads
 \be\label{so-g}
\fg(\xi, t) = 1- e^{-t} + e^{-t} \fg_0(\xi e^t).
 \ee
Reverting to the original variables one obtains the explicit formula
 \be\label{so-f}
 g(x,t) = (1-e^{-t})\, \delta(x=0) + e^{-t}\cdot e^{-t} g_0(x e^{-t}),
 \ee
which shows that the solution at any time $t$ is the convex combination, with precise weights, of the dilated initial density of mean $m_0 e^{t}$ and a Dirac delta function concentrated at zero. Note that the location of the Dirac delta function is uniquely determined by imposing that the growth of the moments of \fer{so-f} is given by \fer{mn}. 

It is remarkable that the average size of the firms is conserved at all finite times $t \ge 0$, so that the limit as $t \to \infty$ of $m_1(t)$ is equal to $m_1(0)$, but $g(x, \infty) = \delta(x=0)$ has vanishing average size. 

This behavior is very close to the situation predicted by the so-called \emph{winner takes all} example in wealth distribution (cf. Chapter $5$ of the book \cite{PT13}). In consequence of the  growth generated by the random variable $\eta_\e$, only one firm will indefinitely increase its size at the expense of the collapse of all the others. 

\section{Diffusion models}\label{diffu}

The example of Section \ref{first} shows that a limiting regime of equation \fer{kin-w} depends on the choice of the random variable $\eta_\e$. At difference with the previous Section, we will  now assume that the random variable $\eta_\e$ is obtained from a centered random variable $X$ taking values on a finite interval $(-1, \gamma)$, where $\gamma >0$,  by multiplication  for the small number $\sqrt \e \ll 1$. Therefore $\eta_\e = \sqrt \e X$. 

The essential difference between the present small perturbation of the size and the previous one of Section \ref{first} is that 
 in the former case any moment of order $n\ge 2$  decays at the same leading order $\e$, while in the latter the  moments of $\eta_\e$ of order $n$ decay at a rate proportional to $\e^{n/2}$. This difference also produces a different equation in the limit.

Let $f_\epsilon(x,\t)$ denote the solution to equation \fer{kin-w} corresponding to the choice $\eta =\eta_\epsilon$. Proceeding as in Section \ref{first}, and denoting $\langle X^2 \rangle = \sigma$,  equation \fer{mome} for $n=2$ gives
 \[
 \frac{d}{d\t}\int_{\R_+}x^2\,f_\epsilon(x,\t)\,dx  = \epsilon \sigma \int_{\R_+} x^2 f_\epsilon(x,\t)\, dx.
 \]
Also in this case, as $\epsilon \to 0$, the second moment of the solution tends to remain constant, thus loosing its increasing property. By scaling time  $t = \epsilon \t $, and denoting $f_\epsilon (x,\t) = u_\epsilon (x, t)$, shows that the second moment of $u_\epsilon$ satisfies the equation
  \be\label{gr-2}
 \frac{d}{dt}\int_{\R_+}x^2\,u_\epsilon(x,t)\,dx  = \sigma \int_{\R_+} x^2 u_\epsilon(x,t)\, dx,
 \ee
which ensures, as in Section \ref{first},  a growth independent of the value of the small parameter $\epsilon$. At difference with the case treated in Section \ref{first}, the limiting equations follows now by expanding $\fu_\e((1+\eta_\e\xi,t)$ in Taylor's series up to the second order. Since $\langle \eta_\e\rangle = 0$, while $\langle \eta_\e^2 \rangle = \e \sigma$, we get
 \be\label{tay}
 \langle \fu_\e((1+\eta_\e\xi,t)\rangle = \fu_\e(\xi,t) + \frac 12 \e \sigma \xi^2 \frac{\partial^2 \fu_\e(\xi,t)}{\partial \xi^2} + \frac 16 \e^{3/2} \xi^3 \left\langle X^3  \frac{\partial^3 \fu_\e(\xi,t)}{\partial \xi^3}\big|_{\xi = \bar\xi}\right\rangle,
 \ee
where $\bar\xi$ is a random number that belongs to the interval $(\xi, 1+ \eta_\e \xi)$. Note that the possibility to expand up to the second order requires the boundedness of the third moment of the initial density. Substituting the expansion into \fer{kin-f} gives
\be\label{kin-u}
 \frac{\partial}{\partial t} \fu_\e(\xi,t) = 
\frac 1\epsilon \left(\frac 12 \e \sigma \xi^2 \frac{\partial^2 \fu_\e(\xi,t)}{\partial \xi^2} + \frac 16 \e^{3/2} \xi^3 \left\langle X^3  \frac{\partial^3 \fu_\e(\xi,t)}{\partial \xi^3}\big|_{\xi = \bar\xi}\right\rangle \right).
 \ee
Hence, letting $\e \to 0$, for any given time $t \ge 0$, at least formally the solution $\fu_\e(\xi, t)$ of equation \fer{kin-u} converges to $\fu(\xi,t)$, solution of the equation
 \be\label{k-u}
 \frac{\partial \fu(\xi,t)}{\partial t}  = \frac \sigma 2 \, \xi^2 \frac{\partial^2 \fu(\xi,t)}{\partial \xi^2}.
 \ee
The limit procedure can be made rigorous by resorting to Fourier-based metrics \cite{PT13}. This result will be proven in details in the Appendix. 

This limit procedure can be clearly done directly resorting to the weak form \fer{kin-w}, by considering smooth functions $\varphi(x)$ of bounded support which additionally satisfy suitable conditions at $x=0$. Indeed, for small values of $\e$, expanding $\varphi(x^*)$ in Tailor's series of $x$ up to the second-order shows that equation \fer{kin-w} for $u_\e$ is well approximated by the equation (in weak form) \cite{CoPaTo05, PT13}
 \be\label{nh}
\frac{d}{dt}\int_{\R_+}\varphi(x)\,u(x,t)\,dx  = 
 \frac \sigma 2 \int_{\R_+} u(x,t) x^2 \varphi''(x)
\,dx. 
 \ee
Integration by parts then shows that equation \fer{nh} coincides with the weak form of the linear diffusion equation \fer{gi-m}, provided the boundary terms produced by integration vanish. Without loss of generality, we will assume $\sigma =2$ in the rest of the paper. In this case, equation \fer{gi-m} reads 
 \be\label{heat-u}
\frac{\partial u}{\partial t} =  \frac{\partial^2}{\partial x^2}(x^2 u). 
 \ee
Note that equation \fer{k-u}, with $\sigma =2$, is the Fourier version of equation \fer{nh}. 

\subsection{The explicit solution of the diffusion equation}
 
For the sake of completeness, we briefly reckon the analytic solution of equation \fer{heat-u}. It is interesting to remark that in a recent paper, Iagar and S\'anchez \cite{IS1} were interested in the study of the asymptotic behavior of solutions to the heat equation in nonhomogeneous media with critical density. The equation for the density $h= h(r, t)$, with  $r \in \R^N$,  $N \ge 3$, and $t \ge 0$, takes the form
\be\label{h-N}
|r|^{-2} \frac{\partial h}{\partial t} = \Delta h. 
\ee
The study of equations of type \fer{h-N} was motivated by a series of papers by Kamin and Rosenau \cite{KR1,KR2,KR3}, devoted to model thermal propagation by radiation in nonhomogeneous plasma. 
As noticed in \cite{IS1}, the results of existence and uniqueness relative to the initial value problem for equation \fer{h-N} with $N \ge 3$ also apply to the one-dimensional problem, which coincides with our equation \fer{k-u} for any $\xi \not= 0$.  

The classical argument in deriving the explicit solution is a suitable transformation of variables, which enables to pass from equation \fer{h-N} to the standard heat equation. In dimension one, this transformation works as follows.
Define 
 \be\label{tra1}
 \fu (\xi, t) = v(y,t); \quad y = \log |\xi| -t.
 \ee 
Then, as $\xi \not=0$
 \[
\frac{\partial \fu}{\partial t} = - \frac{\partial v}{\partial y} + \frac{\partial v}{\partial t},
 \] 
and
 \[
\frac{\partial^2 \fu}{\partial \xi^2} = \frac 1{\xi^2}\left( \frac{\partial^2 v}{\partial y^2} - \frac{\partial v}{\partial y}\right)
 \] 
Hence, if $\fu(\xi,t)$ satisfies equation \fer{k-u}, $v(y,t)$ satisfies the heat equation
 \be\label{heat}
\frac{\partial v}{\partial t} = \frac{\partial^2 v}{\partial y^2}.
 \ee 
It is evident that, by resorting to transformation \fer{tra1}, one can make use of results valid for the heat equation to obtain results for the solution to equation \fer{k-u}.

A similar idea can be used to investigate the diffusion equation \fer{heat-u}. In this case, it is enough to resort, for $x \not= 0$, to the transformation 
 \be\label{tra2}
 u(x,t) = x^{-2}w(x,t); \quad  w(x, t) = v(y,t); \quad y = \log x -t.
 \ee
Then, if $u(x,t)$, with $x \in \R_+$  is a solution to \fer{heat-u}, $w(x,t)$,  $x \in \R_+$ is a solution to  \fer{k-u} (with $\sigma=2$), and finally $v(y,t)$, $y \in \R$ is a solution to the heat equation \fer{heat}. 

In particular, let 
\be\label{max}
M_t(y) = \frac 1{\sqrt{4\pi t}}\exp\left\{ - \frac{y^2}{4t} \right\}
\ee
be the Gaussian probability density of mean zero and variance $2t$, source-type solution of the heat equation \fer{heat} departing from a Dirac delta function located at $y=0$. Then, owing to \fer{tra2} one obtains that the function
 \[
 L_t(x) =  \frac 1{\sqrt{4\pi t}\,x^2}\exp\left\{ - \frac{(\log\, x -t)^2}{4t} \right\}
 \]
is a source-type solution of equation \fer{heat-u}, departing from a Dirac delta function located in $x=1$. In fact, the mean value of $L_t(x)$, for any $t  > 0$ is equal to $1$, since the function $xL_t(x)$, for any $t > 0$ is a lognormal probability density function. Likewise, the second moment of the source-type solution at time $t \ge 0$ is equal to the first moment of the lognormal density $xL_t(x)$. Consequently
 \[
 \int_{\R_+} x^2 L_t(x) \, dx = e^{2t}.
 \]
This implies that the variance of $L_t(x)$ at time $t \ge 0$ is equal to $e^{2t}-1$, and the variance vanishes as $t \to 0$. It is interesting to remark that the source-type solution is itself a lognormal probability density function. This follows from the identity 
 \[
 \frac 1x\,  e^{ - (\log\, x -t)^2/(4t)} = e^{-x}\cdot e^{ - (\log\, x -t)^2/(4t)}= e^{ - (\log\, x +t)^2/(4t)}.
 \]
This proves that the linear diffusion equation \fer{heat-u} possesses a (unique) source-type solution given by the lognormal density 
 \be\label{ln}
 L_t(x) =  \frac 1{\sqrt{4\pi t}\, x}\exp\left\{ - \frac{(\log\, x + t)^2}{4t} \right\}
 \ee
which has been shown to depart at time $t=0$ from a Dirac delta function located in $x=1$.

In analogy with the heat equation \fer{heat}, where the unique solution $v(x,t)$ to the initial value problem is found to be the convolution product of the initial datum $v_0(x)$ with the source-type solution \fer{max}, that is
 \[
 v(y,t) = \int_\R M_t(y-z) v_0(z) \, dz,
 \]
it is a simple exercise to verify that the unique solution to the diffusion equation \fer{heat-u} corresponding to the initial datum $u_0(x)$ is given by the expression
 \be\label{sol}
  u(x,t) = \int_{\R_+} \frac 1z\, u_0\left(\frac xz\right) L_t(z)  \, dz. 
 \ee
It is immediate to show that both the mass and the mean value of the solution \fer{sol} are preserved in time, and the moments of order $n \ge 2$ which are initially bounded increase exponentially at a rate $n(n-1)$. 

\subsection{Large-time behavior}

A further interesting result is concerned with the large-time behavior of the solution to equation \fer{heat-u}. As far as the heat equation \fer{heat} is concerned, it is well-known that the source-type solution \fer{max} represents the intermediate asymptotics of any other solution for a large class of initial data. The recent review article~\cite{Bartier:2011} gives a precise state of the art on this topic.
To make this concept more precise, we define the  {Shannon \emph{entropy}} of a probability density function $f$ as
\begin{equation*}
  \mathcal H(f) := -\int_\R f(z) \, \log f(z) \,  dz.
\end{equation*}
Then it can be shown  (see \eg \cite{Toscani:1996}) that $v(x,t)$ behaves as the source-type solution $\bar M_t$ (the source-type solution with the same variance of $v(x,t)$) when $t \to \infty$, provided that the initial condition $v_0$ is of finite second moment and  entropy.

Moreover, the rate of convergence towards the source-type solution can be computed in $L^1$ norm
\begin{equation} \label{rateCvHeat}
\int_\R | v(z,t) - \bar M_t(z) |\, dz \leq \frac{C}{\sqrt{1+2t}},
\end{equation}
where $C$ is an explicit constant.
The bound \eqref{rateCvHeat} is sharp. A marked improvement of the constant in \fer{rateCvHeat} has been recently obtained in \cite{arnold:2008}, by selecting well parametrized Gaussian functions, characterized either by mass centering or by fixing the second moments or the covariance matrix of the solution.  

The condition of boundedness of the second moment and entropy for the initial value $v_0(y)$ to the heat equation, in view of transformation \fer{tra2} become, for the inital value $u_0(x)$
 \be\label{con1}
 \int_{\R_+} x (\log x)^2 u_0(x) \, dx  < \infty,
 \ee
as far as the second moment of $v_0$ is concerned, and
\be\label{con2}
\left| \int_{\R_+} x\, u_0(x) \log (x^2 u_0(x)) \, dx \right|  < \infty,
 \ee
for the boundedness of entropy. Considering that 
 \[
 \int_{\R_+} x\, u_0(x) \log (x^2 u_0(x)) \, dx =  \int_{\R_+} x (\log x)^2 u_0(x) \, dx + \int_{\R_+} x\, u_0(x) \log u_0(x) \, dx,
 \]
and \fer{con1} guarantees that the first term on the right-hand side is bounded, we can substitute condition \fer{con2} with the following 
 \be\label{ww}
\left| \int_{\R_+} x\, u_0(x) \log  u_0(x) \, dx \right|  < \infty.  
 \ee
Finally, we can rephrase the result about the large-time behavior of the solution to the heat equation for equation \fer{heat-u}. If the initial density  satisfies conditions \fer{con1} and \fer{ww}, the solution to equation \fer{heat-u} converges towards the source-type solution $ \bar L_t$ (the lognormal density with the same mean of $u(t)$), and the following bound holds
 \be\label{conv1}
\int_{\R_+}z | u(z,t) - \bar L_t(z) |\, dz \leq \frac{C}{\sqrt{1+2t}}.
 \ee
Let us observe that, since the mean value of both $u(\cdot, t)$ and $\bar L_t(\cdot)$ is constant, say $ m$, the functions $xu(x,t)/m$ and $x\bar L_t(x)/m$ are probability density functions for all times $t\ge 0$. Consequently, \fer{conv1} is equivalent to the $L^1(\R_+)$ convergence of these probability densities at sharp rate. 

\section{Remarks on the Fourier transform of the lognormal distribution}

Among other applications, the relationship between the kinetic equation \fer{kin-w} and its diffusion approximation \fer{heat-u}, rigorously proven in the Appendix, can be fruitfully used to investigate possible new approximations to the Fourier transform of the lognormal distribution. 
The lognormal distribution is indeed one of the probability distributions most frequently employed in various disciplines which range from physics to chemistry, from engineering to economics, as it arises naturally in a wide variety of applications.  Integral transforms of the lognormal distribution are of great importance in statistics and probability, even if closed-form expressions do not exist. For this reason, a wide variety of methods have been employed to provide approximations, both analytical and numerical (cf. the recent paper \cite{ALR} and the references therein). 
In the absence of a closed-form expression it is clearly desirable to have sharp approximations for the transforms of the lognormal distributions, which are easy to implement numerically.

Thanks to the results of Section \ref{diffu}, and in particular using the fact that the  Fourier transform of the lognormal source-type solution \fer{ln} solves the Fourier transformed version \fer{k-u} of the diffusion equation \fer{heat-u},   one can easily construct a consistent approximation of the Fourier transform of the lognormal distribution, by resorting to the so-called Wild sum representation of the solution to the kinetic equation \fer{kin-w} \cite{PT13}.

In details, let us denote by $F(x)$, $x \ge 0$,  the initial datum of the kinetic equation \fer{kin-w}, in which we fixed $\lambda = 1$ for the sake of simplicity. Moreover, let us suppose that $F(x)$ has enough moments to justify the convergence result of the Appendix. 
Let  $f(\t) \circ M(x)$ denote the gain term in \fer{kin-w}, namely the function such that, for any smooth function $\varphi(x)$
 \be\label{ooo}
\int_{\R^+} \varphi(x) f(\t) \circ M(x) \, dx =   \Big \langle \int_{\R_+} \bigl( \varphi(x^*) \bigr) f(x,\t)
\,dx \Big \rangle,
 \ee
where $x^*$ is given by \fer{col}. Clearly, the symbol $f\circ  M$ stands for the action on $f$ of the random variable $\eta$, distributed  with law $M$. Then,
equation \fer{kin-w} can be fruitfully rewritten 
 in the form
 \be
 \label{wl} \frac{\partial f(x,\t)}{\partial
\t} = f(\t) \circ M(x) - f(x,\t).
  \ee
In this case, one considers the map $f \mapsto \Phi(f)$ given by
 \[
\Phi(f)(\t) = e^{-\t} F + \int_0^\t e^{-(\t-s)}f \circ M\, ds .
 \]
Differentiating on both sides shows that $f(\t)$ solves
the kinetic equation \fer{wl} {exactly when $\Phi(f) = f$}. To find
fixed points one considers iterations. First, put $f^{(0)} = 0$, and
define, for all $j \ge 1$,
 \be\label{ite}
f^{(j+1)}= \Phi\left( f^{(j)} \right).
  \ee
 This yields
 \begin{eqnarray*}
f^{(1)}&=& e^{-t}F\\
f^{(2)}&=& e^{-t}F + t e^{-t}F\circ M\\
f^{(3)}&=& e^{-t}F + t e^{-t}F\circ M + \frac{t^2}2 e^{-t}(F\circ M)
\circ M,
  \end{eqnarray*}
   and so on.
Clearly
 \[
f^{(j+1)}- f^{(j)} \ge 0
  \]
for all $j \ge 1$.   The function
 \[
 f(x,\t)= \lim_{j \to \infty}f^{(j)}(\t),
  \]
the limit of the monotone sequence of the $f_j(t)$,  exists, and it is a
solution to the kinetic equation \fer{wl}. Note that
 \be\label{wwl}
f(x,\t) = e^{-\t} \sum_{k=0}^\infty \frac{\t^k}{k!} f^{(k+1)}(x),
 \ee
where the positive coefficients $f^{(k)}$, $k \ge 1$, are
recursively defined by
 \[
f^{(k+1)} = f^{(k)}\circ M,
 \]
starting from $f^{(1)}= F$. It is important to remark that, at any
time $\t \ge 0$, $f(x,\t)$, as given by \fer{wwl}, is a convex
combination of the (time-independent) coefficients $f^{(k)}$.

Historically, the idea of introducing an increasing sequence to solve kinetic equations is due to  Wild~\cite{Wild}  who proved by this idea the existence of
solutions to the Boltzmann equation for Maxwell molecules. The argument of Wild  was
completed by Morgenstern~\cite{Mor}, who proved the uniqueness of
solutions to the same equation three
years later. Let
us discuss briefly the importance of Wild's argument. His idea
immediately leads to the construction of a monotone sequence which
approximates the solution, in which the approximations are made by subsequent
iterations. Hence, the Wild approximation enters deeply into
the structure of the solution to the  kinetic equation. This
idea has been developed in a number of papers in which the
approximation has been clarified for the Kac model~\cite{McK2} from a
probabilistic point of view.

Formula \fer{wwl} immediately gives the expression of the solution in Wild sum for the Fourier transformed version of the kinetic equation. Given the initial value $\widehat F(\xi)$ of equation \fer{kin-f} (with $\lambda =1$), the solution $\ff(\xi)$ can be expressed in the form

 \be\label{wwf}
\ff(\xi,\t) = e^{-\t} \sum_{k=0}^\infty \frac{\t^k}{k!} \ff^{(k+1)}(\xi),
 \ee
where the  coefficients $\ff^{(k)}$, $k \ge 1$, are
recursively defined by
 \[
\ff^{(k+1)} = \langle \ff^{(k)}((1+\eta)\xi)\rangle,
 \]
starting from $\ff^{(1)}= \widehat F$.

Formula \fer{wwf} can be easily adapted to give an expression of the solution to equation 
\be\label{kin-uu}
 \frac{\partial}{\partial t} \ff_\e(\xi,t) = 
\frac 1\epsilon \left( \langle \ff_\e((1+\eta_\epsilon)\xi, t)\rangle -\ff_\e(\xi,t) \right), 
 \ee
which approximates, for $\epsilon \ll 1$, the solution to the diffusion equation \fer{heat-u}.
For example, one can  set  $\eta_\epsilon$ to be a two-valued random variable that takes the values $-\sqrt{2\epsilon}$ and $+\sqrt{2\epsilon}$ with probability $1/2$, and as initial value the function $\ff^{(1)}(\xi)= \widehat F(\xi) = e^{-i\xi}$,  the Fourier transform of a Dirac delta function located in $x=1$. Then $\langle \eta_\epsilon \rangle= 0$, $\langle \eta_\epsilon^2 \rangle= 2$, and the  coefficients $\ff^{(k)}$, $k \ge 1$, are
recursively defined by
 \[
\ff^{(k+1)} = \frac 12 \left( \ff^{(k)}((1-\sqrt{2\epsilon})\xi)+ \ff^{(k)}((1+\sqrt{2\epsilon})\xi)\right),
 \]
 which give, for  the Fourier transform of the lognormal density \fer{ln} the approximate expression
 \be\label{ln-f}
 \widehat L_t(\xi) \cong e^{-t/\epsilon} \sum_{k=0}^\infty \frac{(t/\epsilon)^k}{k!} \ff^{(k+1)}(\xi), \qquad \epsilon \ll 1.
 \ee
Formula \fer{ln-f} is very easy to implement, and it will be dealt with in a companion paper. We refer to \cite{PR00, PR01, PR01b} for some recent applications of Wild sum's to the solution of the Boltzmann equation.

\section{Conclusions}

In this paper, we investigated the possible continuous limit equations that can be obtained from the classical law of proportionate effect proposed by Gibrat \cite{Gib1, Gib2} to justify the observed distribution of firms in a society. Among others, the diffusion limit of the underlying kinetic model constructed by following Gibrat's is highly interesting, and it is deeply connected to the diffusion equation for the geometric Brownian motion \cite{Oks}. The limit equation is rigorously derived as soon as the initial value possesses a certain number of moments bounded. 

This rigorous limit allows to use the expression of the solution to the kinetic model, which is fruitfully expressed in a way easily computable by a recursive argument, to obtain an approximate expression of the Fourier transform of a lognormal density. 

\bigskip\noindent{ \bf{\emph{Acknowledgments.}}} { \small
 Support by MIUR
project ``Optimal mass transportation, geometrical and functional
inequalities with applications'' and by the National Group of
Mathematical Physics of INDAM is kindly acknowlwedged. }

\eject

\section{Appendix}
In this appendix, we will 
justify the limiting behaviour (as $\e \to 0$) of the solution to equation
\fer{kin-u}. We refer to \cite{MaTo07, PT13} for further details. Convergence will be proven in terms  of Fourier based metrics \cite{Carrillo:2007, PT13}.  Given $s >0$ and two probability densities $f$ and $g$, their Fourier based distance $d_s(f,g)$ is the quantity
            \begin{equation*}
            d_s(f,g) := \sup_{\xi \in \RR \setminus 0} \frac{\left |\widehat{f}(\xi)-\widehat{g}(\xi)\right |}{|\xi|^s}.
            \end{equation*}
            This distance is finite, provided that $f$ and $f$ have the same moments up to order $[s]$, where, if $s \notin \NN$,  $[s]$ denotes the entire part of $s$, or up to order $s-1$ if $s \in \NN$. Moreover $d_s$ is an \emph{ideal} metric, equivalent to the weak*-convergence of measures \cite{Carrillo:2007}.

For reasons of simplicity, and to highlight the role of $\e$, we set in \fer{kin-f} $\lambda =1$,  $t=\e \t$, and $f(x,\t) = u_\e(x,t)$. In the Fourier transform,
equation \fer{kin-f} then takes the form
 \be\label{e-dri}
\frac{\partial \fu_\e(\xi,t) }{\partial t}  = \frac 1\e \left(
\fu_\e((1+\eta_\e)\xi,t) - \fu_\e(\xi,t) \right).
 \ee
If the common initial value of equations \fer{k-u} and \fer{e-dri} is such that the third moment of the solution is  bounded, the
relationship between equations \fer{k-u} and  \fer{e-dri} can be outlined immediately.
 In fact, thanks to the Tailor expansions \fer{tay}
 we can use the identity
\[
\frac 12 \sigma \xi^2 \frac{\partial^2 \fu(\xi,t)}{\partial \xi^2} = \frac 1\e \left(\langle \fu((1+\eta_\e\xi,t)\rangle - \fu(\xi,t)\right) - \frac 16 \e^{1/2} \xi^3 \left\langle X^3  \frac{\partial^3 \fu(\xi,t)}{\partial \xi^3}\big|_{\xi = \bar\xi}\right\rangle,
\]
valid for a suitable $\bar\xi$,  random number that belongs to the interval $(\xi, 1+ \eta_\e \xi)$.
Hence, equation \fer{k-u} for $\fu$ can be written as
  \be\label{ee-dri}
\frac{\partial \fu(\xi,t) }{\partial t}  = \frac 1\e \left( \langle
\fu((1+\eta_\e)\xi,t) \rangle- \fu(\xi,t) \right) - \e^{1/2}\,\xi^3 R(\xi, t),
 \ee
 where the remainder term $R(\xi, t)$ is given by
  \[
 R(\xi, t)= 
 \frac 16  \left\langle X^3  \frac{\partial^3 \fu(\xi,t)}{\partial \xi^3}\big|_{\xi = \bar\xi}\right\rangle
 \]
Equations \fer{e-dri} and \fer{ee-dri} differ only by the presence of a
term of size proportional to $\e^{1/2}$ (the last term in \fer{ee-dri}). We
remark that, by construction, the solutions to both equations are such
that  mass and momentum are preserved (equal to one), while the second moment has the
same law of growth, independent of $\e$, as given by \fer{gr-2}. Hence, by choosing a common initial value in both equations with bounded moments up to order three,  the two solutions have
the same moments up to the order two at any subsequent time. This implies that  the
Fourier-based metric $d_3$ of the solutions is bounded in time if it is bounded initially. Bearing this in mind, for
any given $\xi \not=0$, let us subtract equation \fer{ee-dri} from
equation \fer{e-dri}, and divide both sides by $|\xi|^3$ to obtain
 \begin{eqnarray}
 \nonumber
\frac{\partial }{\partial t}\frac{\fu_\e(\xi,t) -\fu(\xi,t)}{|\xi|^3}+ \frac 1\e \frac{\fu_\e(\xi,t) -\fu(\xi,t)}{|\xi|^3} = \qquad\qquad
\\[-.25cm]
 \label{equ33}
 \\[-.25cm]
 \nonumber
\frac 1\e \frac{\langle \fu_\e((1+\eta_\e)\xi,t) -\fu((1+\eta_\e)\xi,t)\rangle}{|\xi|^3} + \e^{1/2}\,\frac{\xi^3}{|\xi|^3} R(\xi, t)
.
 \end{eqnarray}
Now, consider that the solution to  equation \fer{gi-m} satisfies 
 \[
\int_{\R_+} |x|^3u(x,t) \, dx = e^{6\sigma t}\int_{\Re_+}|x|^3 u(x,t=0) \, dx = m_3 e^{6\sigma t},
 \]
 where we denoted by $m_3$ the third principal moment of the initial value. 
Therefore, by a classical property of Fourier transforms,
 \[
\left|\frac{\xi^3}{|\xi|^3} R(\xi, t) \right| \le |\langle X^3 \rangle | m_3 e^{6\sigma t},
 \]

Using this upper bound in \fer{equ33} ,  by setting
 \[
h_\e(\xi,t) = \frac{\fu_\e(\xi,t) -u(\xi,t)}{|\xi|^3},
 \]
we obtain that $h_\e(t)$ satisfies the inequality
 \[
\frac{\partial h_\e(\xi,t)}{\partial t} + \frac 1\e
h_\e(\xi,t)\le \frac 1\e \sup_\xi  \frac{|\langle \fu_\e((1+\eta_\e)\xi,t) -\fu((1+\eta_\e)\xi,t)\rangle |}{|\xi|^3} + \e^{1/2}\, m_3 e^{6\sigma t}
 \]
 \[
\le  \frac 1\e \langle (1+\eta_\e)^3 \rangle \|h_\e \|_\infty(t)  + \e^{1/2}\, m_3 e^{6\sigma t}. 
 \]
Note that
 \[
 \langle (1+\eta_\e)^3 \rangle = 1 + 3\sigma \e + \langle X^3\rangle \e^{3/2} = c(\e).
 \]
Hence, the previous inequality is equivalent to
 \[
 \frac{\partial}{\partial t} \left (h(\xi,t) e^{ t/\e } \right ) \leq
\frac {c(\e)}\e \|h(\cdot, t)  e^{t/\e }\|_\infty  +   \e^{1/2}\, m_3 e^{(6\sigma + 1/\e) t}.
 \]
 Integrating from 0 to $t$, we get
 \[
  h(\xi,t) e^{ t/\e } \leq h(\xi,0) + \int_0^t \e^{1/2}\, m_3 e^{(6\sigma + 1/\e) s}\, ds + \int_0^t \frac {c(\e)}\e \|h(\cdot, s)  e^{s/\e }\|_\infty \, ds.
 \]
 Hence, if $H(t) =\|h(\cdot, t) e^{ t/\e }\|_\infty$, and
 \be\label{gron}
 \psi(t) = H(0) + \int_0^t  \e^{1/2}\, m_3 e^{(6\sigma + 1/\e) s}\, ds,
 \ee
 $H(t)$ satisfies
 \[
 H(t) \leq \psi(t) + \int_0^t \frac  {c(\e)}\e
 H(s)\,ds.
\]
 Now, by the generalized Gronwall inequality, %\index{Gronwall inequality}
\[ 
\kappa(t) \leq \psi(t) + \int_0^t \lambda(s) \kappa(s)\, ds
 \]
implies
\[
 \kappa(t) \leq \psi(0) \exp\left \{ \int_0^t \lambda(s)\, ds\right \}
+ \int_0^t \exp \left\{\int_s^t \lambda(r)\, dr \right \}
\frac{d\psi}{ds}\, ds.
 \]
Applying this inequality with $\lambda(t) = c(\e)/\e$ and
$\psi(t)$ given by \fer{gron} we obtain
 \[
H(t) \le \left[ H(0) + \e^{1/2} A_\e(t)\right] e^{(c(\e)/\e)t},
 \]
where
 \[
A_\e(t) = \frac {m_3}{ 3\sigma - \langle X^3\rangle \e^{1/2}} \left(\exp\{ (  3\sigma - \langle X^3\rangle \e^{1/2})t\}- 1 \right). 
 \] 
Note that the denominator of the previous expression is positive for sufficiently small $\e$. 
Going back to $h_\e(\cdot,t)$, we finally obtain
 \[
\|h(\cdot, t) \|_\infty \le \left[ \|h(\cdot, 0) \|_\infty  + \e^{1/2} A_\e(t) \right]\exp\{ (  3\sigma + \langle X^3\rangle \e^{1/2})t\} .
 \]
Recalling now that $\|h(\cdot, t) \|_\infty =d_3(u_\e, u)(t)$, we conclude
with the bound
 \be\label{un-bo}
d_3(u_\e, u)(t) \le \left[ d_3(u_\e, u)(0)  + \e^{1/2} A_\e(t)
\right] \exp\{ (  3\sigma + \langle X^3\rangle \e^{1/2})t\},
 \ee
which holds uniformly with respect to $\e$ and $t$, provided the
distance between the initial data is finite. In particular,
by taking the same initial density for both equation
\fer{e-dri} and the kinetic equation \fer{ee-dri} one concludes with
the bound
 \be\label{un-bo2}
d_3(u_\e, u)(t) \le  \e^{1/2} A_\e(t)\exp\{ (  3\sigma + \langle X^3\rangle \e^{1/2})t\},
 \ee
which proves that as soon as $\e \to 0$ the solution to the kinetic
equation converges towards the solution to  equation \fer{k-u} for any 
time $t >0$.

%%%%%%%%%%%%%%%%%%%%%%%%%%%%%%%%%%%%%%%%%%%%%%%%%%%%%%%%%%

\end{document}